# Comparison of extended irreversible thermodynamics and nonequilibrium statistical operator method with thermodynamics based on a distribution containing the first-passage time

V. V. Ryazanov

*Institute for Nuclear Research, pr. Nauki, 47 Kiev, Ukraine, e-mail: vryazan19@gmail.com*

An analogy is drawn between version of non-equilibrium thermodynamics a distribution-based containing an additional thermodynamic first-passage time parameter, nonequilibrium statistical operator method and extended irreversible thermodynamics with flows as an additional thermodynamic parameter. Thermodynamics containing an additional thermodynamic first-passage time parameter maps to extended irreversible thermodynamics. Various conditions for the dependence of the distribution parameters of the first-passage time on the random value of energy, the first thermodynamic parameter, are considered. Time parameter relaxation time $\tau$ of extended irreversible thermodynamics is replaced by the average first-passage time. Expressions are obtained for the thermodynamic parameter, the conjugate of the first passage time through the entropy change, and for the average first passage time through the flows.
    Keywords: Extended Irreversible Thermodynamics, first-passage time, entropy change and flows

## 1. Introduction

In non-equilibrium thermodynamics, in contrast to equilibrium, there are several different directions. That is classical irreversible thermodynamics [1-4], this direction is described in detail in [5]; generalization of classical irreversible thermodynamics to states far from equilibrium and to nonlinear phenomena [6]; rational thermodynamics [7]; extended irreversible thermodynamics [8-13]; generalized kinetic approach [14]; wave approach to thermodynamics [15]; "universal" approach [16]; stochastic thermodynamics [17]; information statistical thermodynamics [18], [19-33], and etc. This brief history of nonequilibrium thermodynamics is described in more detail in [20], [77].

The method of non-equilibrium statistical operator (*NSO*) [34-43] turned out to be very effective, on the basis of which informational statistical thermodynamics [19-33] was developed. It was shown in [44] that the *NSO* operator is represented as a quasi-equilibrium operator averaged over the distribution of the past lifetime of the system.

In [45-48], a statistical distribution is introduced containing a random lifetime or a random time when a random process reaches a certain level, first-passage time (*FPT*). *FPT* is studied in detail and widely and effectively used in various problems [49–68]. Based on a distribution containing *FPT*, it is possible to develop nonequilibrium thermodynamics with a nonequilibrium thermodynamic parameter *FPT* [69–70]. *FPT* or lifetime is also contained in the *NSO* method [44]. An additional thermodynamic parameter responsible for the nonequilibrium description is *FPT*. An additional thermodynamic parameter is also contained in extended irreversible thermodynamics (*EIT*) [8–13]. In *EIT*, these are flows, a value close to *FPT*.

The use of an additional nonlocal flow parameter in extended irreversible thermodynamics [8–13] does not allow using the local equilibrium hypothesis. Such classical regularities as Fourier's and Fick's laws are generalized. The theory includes memory effects, non-local and nonlinear effects. Generalized entropy and entropy production rate are at the center of consideration. Entropy is represented as a function not only of conserved variables, but also of dissipative flows. Flows are specified as state variables (see [8-13]).

In informational statistical thermodynamics [20–33], the fluxes and relaxation times play an important role. This article compares thermodynamics with thermodynamic variable *FPT* (*TFPT*) with the *NSO* method and *EIT*.

The article contains several original points. Thus, explicit expressions are obtained for the conjugate *FPT* of the thermodynamic parameter $\gamma$ in terms of the change in entropy. The parameter $\gamma$ (11) is similar to the reciprocal temperature $\beta$, which in the statistical Gibbs distribution is conjugate to the thermodynamic random energy parameter $u$. Accordingly, the definition of the parameter $\gamma$ is similar to the definition of the reciprocal temperature, this is an important point. The average *FPT* values depend on the parameter $\gamma$. Substituting certain values of the parameter $\gamma$ leads to expressions of the mean *FPT*s in terms of entropy changes and fluxes. Inverse expressions are also obtained: flows through the average *FPT*. Another new point is the generalization of the relaxation time $\tau$ to *EIT*, replacing $\tau$ with $\bar{\tau}(\beta,\gamma)$, the mean *FPT*. The article also considers models for the dependence of the *FPT* distribution on the thermodynamic value of random energy $u$. Such dependencies significantly increase the possibilities of statistical description.

The article is structured as follows. In the second section, the distribution of the *NSO* in the case of the classical description is equated to the distribution containing the *FPT*. An explicit expression is obtained for the thermodynamic parameter conjugate to *FPT*, an analogue of the reciprocal temperature, conjugate random energy, depending on the change in entropy. Then this expression is obtained in a general form. Section 3 compares *EIT* and *TFPT*. Relaxation time is generalized and acts as *FPT*. In conclusion, the results obtained are discussed.

## 2. NSO and TFPT

### 2.1. NSO and FPT

One of the most general in modern nonequilibrium statistical mechanics is the NSO method. It can serve as the basis for other important approaches based on Mori and Kawasaki–Ganton projection operators. [43]. The approach of Zubarev to the theory of non-equilibrium processes [34-41] based on a fundamental property of macroscopic systems, which is intimately related to phase space $z=(q(t), p(t))$ of a many-particle system. For time intervals that are not too small in the instability of the classical phase trajectories $z(t)$, the details of the initial state become unimportant and the number of parameters necessary for the description of the state of the system is reduced.

It was shown in [44] that the *NSO* can be written as

$$\ln \rho_w(t) = \int_0^\infty w(y) \ln \rho_{rel}(t-y,-y) dy = \ln \rho_{rel}(t,0) - \int_0^\infty (\int dy w(y)) \frac{d \ln \rho_{rel}(t-y,-y)}{dy} dy, \quad (1)$$

where $y=t-t_0$, $t$ is the current moment of time and $t_0$ is the random initial moment. Averaging the logarithm of the relevant distribution $\rho_{rel}$ over the distribution $w(y)$ of the past lifetime of the system $y=t-t_0$ leads to the logarithm *NSO* (1). The logarithm *NSO* $\rho_w(t)$ expression (1) is the logarithm of the relevant distribution $\rho_{rel}(t,0)$ (2) averaged over the distribution of the lifetime $w(y)$. Thus, the history of the system is taken into account.

In [38] another physical interpretation of expression (1) is given: the system described by the logarithm of the relevant distribution is isolated. But at random times defined by the exponential distribution (for case (6)) or, more generally, by the distribution $w(y)$, it falls under the influence of the environment and makes random transitions.

The non-equilibrium macroscopic state of the system depends on the set of observables, average values $<\hat{P}_m>$ of the corresponding dynamic variables $\hat{P}_m$. Variables $\hat{P}_m$ characterize the reduced description of the state of the system. The choice of the set of variables $\hat{P}_m$ is arbitrary. It is determined by the specifics of the task and the stage of evolution at which the system is located. The stages of evolution, the chosen time scales, depend on the "hierarchy" of the basic relaxation times in macroscopic systems [20, 23, 34, 35]. An increase in the evolution time of the system is accompanied by a decrease in the variables necessary to describe the behavior of the system [35]. In this article, relaxation times are interpreted as *FPT*.

In the case when the observed macroscopic quantities depend on time, generalized Gibbs ensembles are constructed for the thermodynamic description of nonequilibrium systems [35]. They are called relevant ensembles. Irreversible macroscopic processes are described by solutions of the Liouville equation [20-33]. These the observed variables appear in auxiliary distributions. To select the basic set of variables, the Zubarev-Peletminskii law [20] is used. Using the information entropy maximum method and Lagrange multipliers in [20], [34], [35], the relevant [35] (or quasi-equilibrium [38]) distribution is obtained in the form

$$\rho_{rel}(t,0) = \exp[-\Phi(t) - \sum_m F_m(t)\hat{P}_m]. \qquad (2)$$

A set of variables $\{F_m(t)\}$ play the role of variables thermodynamically conjugated to the macroscopical variables $\hat{P}_m$. As in the Gibbs distribution, the reciprocal temperature $\beta$ is conjugated to random energy $u$. The first $t$ in the argument of $\rho_{rel}(t,0)$ denotes the time dependence of the thermodynamic variables $F_m(t)$; the second temporal argument refers to the Schrödinger representation of dynamical operators and their evolution [34, 35],

$$\rho_{rel}(t',t'-t) = \exp[-i(t'-t)L]\rho_{rel}(t',0).$$

Here $iL\rho = [H,\rho] = \sum_k [\frac{\partial H}{\partial q_k}\frac{\partial \rho}{\partial p_k} - \frac{\partial H}{\partial p_k}\frac{\partial \rho}{\partial q_k}]$, $L$ is Liouville operator; $H$ is Hamilton function, $p_k$ and $q_k$ are pulses and coordinates of particles; [...] is Poisson bracket [34, 35].

The function $\Phi(t)$ from (2) is determined from the normalization condition of the distribution (2) and has the form

$$\Phi(t) = \ln Tr(\exp\{-\sum_m F_m(t)\hat{P}_m\}).$$

From the self-consistency conditions are determined the Lagrange multipliers $F_m(t)$ [34, 35, 41]

$$\left\langle \hat{P}_m \right\rangle^t = \left\langle \hat{P}_m \right\rangle^t_{rel} = Tr(\hat{P}_m \rho_{rel}(t,0)), \quad Tr\rho_{rel}(t,0) = 1.$$

The considerations are concerned with classical and quantum systems if the dynamical variables and the symbol *Tr* are correspondingly interpreted.

When obtaining an explicit form of a non-equilibrium statistical operator, the principle of maximizing the statistical-informational entropy is used under additional conditions that are set at each moment of the system's past [34]. This turns on the memory effect.

In [20, 23, 34] is maximized the time-dependent Gibbs' entropy

$$s_G(t) = -Tr\{\rho_w(t)\ln \rho_w(t)\}$$

under additional conditions

$$Tr\{\rho_w(t`)\} = 1, \quad Q_m(r,t`) = Tr(\hat{P}_m \rho_w(t`)) = Tr(\hat{P}_m \rho_{rel}(t`)) \tag{3}$$

for $t_0 \leq t' \leq t$ [20, 34]. In (3), $Q_m(r,t`)$ values are equal to the average $<\hat{P}_m>$ of the corresponding dynamic variables $\hat{P}_m$. This procedure reflects the history of the system. The result is

$$\rho_w(t) = \exp\{-\Psi(t) - \sum_{j=1}^{n}\int_{t_0}^{t} dt' \int d^3 r \varphi(r,t,t')\hat{P}_j(r)\}, \tag{4}$$

where

$$\Psi(t) = \ln Tr\{\exp[-\sum_{j=1}^{n}\int_{t_0}^{t} dt' \int d^3 r \varphi(r,t,t')\hat{P}_j(r)]\}$$

corresponds to the normalization $\rho_w(t)$. From relations (3) the Lagrange multipliers are found. They are redefined as

$$\varphi_j(r,t,t') = w(t,t')F_j(r,t'), \tag{5}$$

functions $w$ is present here [20, 23]. They are introduced by this relation [34].

In [34, 35], the volume of the system tends to infinity (thermodynamic limit), then the limit transition is made $t_0 \to -\infty$, $t_0-t \to -\infty$. Before this passage to the limit, the functions in (4) depend on the value of $y=t-t_0$. The value $y=t-t_0$ in [44] is considered as a random system lifetime, *FPT* (in the reverse time). The functions $w(t,t')$ are interpreted in [44] as the densities of the distribution of this lifetime.

Zubarev in [34] chose the weight function $w(t, t')$ (or the distribution density in treatment [44]) in expression (5) in the form of the Abel kernel $\varepsilon \exp\{\varepsilon(t'-t)\}$ (or the exponential distribution [44]). To eliminate non-physical initial correlations at the initial moment $t_0$, the limit transition $t_0 \to -\infty$, $t_0-t \to -\infty$ is made. Hence,

$$\rho_\varepsilon(t) = \exp\{-\varepsilon \int_{-\infty}^{t} dt' e^{\varepsilon(t'-t)} \hat{S}(t', t'-t)\}. \tag{6}$$

Eq. (6) after part integration takes the form

$$\rho_\varepsilon(t) = \exp\{-\hat{S}(t,0) + \hat{\zeta}_\varepsilon(t)\}, \quad \hat{\zeta}_\varepsilon(t) = \int_{-\infty}^{t} dt' e^{\varepsilon(t'-t)} \frac{d}{dt'}\hat{S}(t', t'-t), \tag{7}$$

where $\rho_{rel}(t,0) = \exp\{-\hat{S}(t,0)\}$, $\hat{S}(t,0) = \Phi(t)\hat{1} + \sum_{j=1}^{n}\int d^3 r F_j(r,t)\hat{P}_j(r)$ is entropy operator.

The entropy production operator [34, 35] is equal

$$\frac{d\ln\rho_{rel}(t-y,-y)}{dy} = -\hat{\sigma}(t-y,-y) = -\frac{d\hat{S}(t-y,-y)}{dy}; \quad \frac{d\hat{S}(t)}{dt} = \sum_m \frac{\partial F_m(t)}{\partial t}(\hat{P}_m - \langle\hat{P}_m\rangle^t) + \sum_m F_m(t)\dot{\hat{P}}_m. \tag{8}$$

The expression for entropy production $\hat{\sigma}$, $\bar{\sigma}$ for the local-equilibrium distribution, a special case of the relevant distribution (2), is written in terms of flows in [34]. Further more detailed study of these relations, as well as expressions for flows, was carried out in [20-33].

In *NSO* (6), the distribution $w(y)$ is chosen as an exponential distribution valid for long times. After the limit transition $t_0 \to -\infty$, explicit dependence on $y$ in (4) passes into averaging over $y$. *NSO* becomes an integrated function over $y$, we arrive at averaging, as in (6) and (1) (after a change of variables).

How *FPT* is determined. At [45] it is written: «*FPT* is the time it takes for a stochastic process $X(t)$ to reach a certain threshold $a$ (9) for the first time. For example, it is the time for a stochastic process that describes a macroscopic parameter $X(t)$ to reach the zero level. The stochastic process *X(t)* is the order parameter of the system, for example, its energy, number of particles, (in some papers, for example [71], entropy production is considered). The process *X(t)* can describe many other physical quantities. *FPT* is, by definition,

$$T_{\gamma x} = \inf\{t : X(t) = a\}, \quad X(0) = x > 0. \tag{9}$$

The subscript $\gamma$, emphasizing the dependence on the conjugate thermodynamic parameter, is used not to confuse the variable with the temperature *T*. There are other definitions of *FPT*. For example, $T_{\gamma x} = \inf\{t > 0; X(t) \in A\}$, where *A* is a Borel set on a number line, $T_{\gamma x}$ is the moment of the first achievement of the set *A* [72]. The moments of the first achievement of the *FPT* refer to the stopping times and to the Markov moments [72, 73]. Each Markov moment is associated with a set of sets describing a set of events observed over a random time $T_{\gamma x}$ [73]. Physically, this corresponds to a dependency on the system's history. The events that occur during this time include the change in the entropy of the system. In this way, the system history is taken into account. *FPT* (9) is a multiplicative functional from a random process *X(t)* [72]. In distributions (11)-(13), as in the *NSO* method, the dependence on the system's past is important.

By definition (9) *FPT* is the first moment when a random process *X(t)* reaches the absorbing set "*a*". For example, in relation (11), the role of the main random process *X(t)* is played by the internal energy *u*. In example, this is a random flux *J*; this is the coordinate of a diffusing particle. *FPT* $T_\gamma$ plays the role of a subordinate process [74]. In *NSO* (1), (6), (7) the absorbing set "*a = 0*" is achieved in the reverse time».

Non-physical dependence on the initial moment $t_0$ is eliminated by choosing a sufficiently large interval $t$-$t_0$ so that non-physical initial states do not influence. In this case, the time $t$-$t_0$ should not exceed the time to relaxation to equilibrium of the entire system.

In the stationary case, when the entropy production operator does not depend on time, in the second term on the right side of expression (1), the entropy production operator $\hat{\sigma}_{st}$ is taken out of the integration, and expression (1) takes the form

$$ln\rho_w(t) = ln\rho_{rel}(t,0) - <t-t_0> \hat{\sigma}_{st}. \tag{10}$$

In expression (10), there may be not stationary $\hat{\sigma}_{st}$, but $\hat{\sigma}(t-y,-y) \simeq \hat{\sigma}(t)$, a function that does not depend on the past $y=t$-$t_0$.

### 2.2. TFPT and NSO. Relationship between the parameter γ and entropy.

In [45–48], a statistical distribution is introduced that contains an additional thermodynamic parameter *FPT* $T_\gamma$ (or the lifetime)

$$\rho(z;u,T_\gamma) = \exp\{-\beta u - \gamma T_\gamma\} / Z(\beta,\gamma). \tag{11}$$

From the microscopic (coarse-grained) density $\rho(z;u,T_\gamma)$ (11) we pass to the distribution density of random variables of energy *u* and $T_\gamma$ $p(u,T_\gamma)=p_{uT\gamma}(x,y)$ (for example, see [75], [76]):

$$p(u,T_\gamma) = \int \delta(u-u(z))\delta(T_\gamma - T_\gamma(z))\rho(z;u(z)T_\gamma(z))dz. \tag{12}$$

After passing from variables *z* to variables *u*, $T_\gamma$, substituting expression (11) into relation (12), and changing variables, we obtain

$$p(u,T_\gamma) = \frac{e^{-\beta u - \gamma T_\gamma} \omega(u,T_\gamma)}{Z(\beta,\gamma)}, \tag{13}$$

where $\beta = 1/T$ is the inverse temperature of the reservoir ($k_B = 1$, $k_B$ is Boltzmann constant),

$$Z(\beta,\gamma) = \int e^{-\beta u - \gamma T_\gamma} dz = \iint du \, dT_\gamma \, \omega(u,T_\gamma) e^{-\beta u - \gamma T_\gamma} \tag{14}$$

is the partition function. The Lagrange multipliers, parameters $\beta$ and $\gamma$ can be expressed in terms of average values $\bar{u}$, $\bar{T}_\gamma$ from relations (3). Differentiating $\ln Z(\beta,\gamma)$ with respect $\beta$ and $\gamma$, we obtain the thermodynamic equalities:

$$\langle u \rangle = -\frac{\partial \ln Z(\beta,\gamma)}{\partial \beta}\bigg|_\gamma, \qquad \langle T_\gamma \rangle = -\frac{\partial \ln Z(\beta,\gamma)}{\partial \gamma}\bigg|_\beta. \tag{15}$$

The function $\omega(u,T_\gamma)$ is the volume of the hypersurface in the phase space. It contains fixed values of $u$ and $T_\gamma$. The function $\omega(u,T_\gamma)$ for the variables $u$ and $T_\gamma$ replaces the factor $\omega(u)$ in the case of distribution for one variable $u$. Function $\omega(u,T_\gamma)=d^2\mu(u,T_\gamma)/dudT_\gamma$, where $\mu(u,T_\gamma)$ is the number of states in the phase space with parameter values less than $u$ and $T_\gamma$. In this case, $\int\omega(u,T_\gamma)dT_\gamma=\omega(u)$. The function $\omega(u,T_\gamma)dudT_\gamma$ is equal to the number of phase points with parameters in the interval between $u$, $u+du$; $T_\gamma$, $T_\gamma+dT_\gamma$.

The joint probability for $u$ and $T_\gamma$ is (13), (14). The function $\omega(u,T_\gamma)$ is equal to the joint probability for $u$ and $T_\gamma$. The important thing is that $\omega(u,T_\gamma)$ is the stationary probability of the process for $u$ and $T_\gamma$. Let's rewrite the value $\omega(u,T_\gamma)$ in the form

$$\omega(u,T_\gamma) = \omega(u)\omega_1(u,T_\gamma) = \omega(u)\sum_{k=1}^n R_k f_{1k}(T_\gamma,u). \tag{16}$$

Relation (16) assumes that there are $n$ classes of states in the system. Below we restrict ourselves to the case $n=1$. The value $R_k$ is equal to the probability that the system is in the $k$-th class of states; the function $f_{1k}(T_\gamma,u)$ is equal to the *FPT* $T_\gamma$ distribution density in this class of (ergodic) states (in the general case, $f_{1k}(T_\gamma,u)$ depends on $u$). Such functions (16) characterize the behavior of metals, glasses, etc.

The dependence on the argument random energy $u$ in expression (11) is conditional and can be changed to a more general expression using expression (2),

$$\rho(z;u,T_\gamma) = \rho_{rel}Z_\beta \exp\{-\gamma T_\gamma\}/Z(\beta,\gamma), \qquad Z_\beta = \exp[\Phi]. \tag{17}$$

As in *NSO*, the distribution (2) is generalized. In (17) contains an additional thermodynamic parameter *FPT* $T_\gamma$ in addition to the thermodynamic parameters $\hat{P}_m$ from (2).

If we assume the possibility of instrumental measurement of the *FPT*, then we can introduce the specific entropy $s_\gamma$ corresponding to distribution (11) ($u$ is the specific internal energy) by the relation [34–35]

$$s_\gamma = -\langle \ln \rho(z;u,T_\gamma)\rangle = \beta\langle u\rangle + \gamma\langle T_\gamma\rangle + \ln Z(\beta,\gamma); \qquad ds_\gamma = \beta d\langle u\rangle + \gamma d\langle T_\gamma\rangle. \tag{18}$$

Let us average expression (10) over distribution (1). At the same time, we take into account that [20, 23, 25-27, 34-35] $-\int dz\rho_w(z,t)\ln\rho_w(z,t) = s_G$, Gibbs' statistical entropy [23]. We suppose $s_G = s_\beta$ (21). To the entropy $-\int dz\rho_w(z,t)\ln\rho_{rel}(z;t,0) = -\int dz\rho_{rel}(z;t,0)\ln\rho_{rel}(z;t,0) = \bar{s}(t)$ [23, 25-27, 34-35] we equate the entropy $s_\gamma$ (18). Then averaging (10) over distribution (1) gives the relation

$$<t-t_0>\bar{\sigma}_{st} = s_\beta - \bar{s} = \Delta s, \tag{19}$$

where $\int dz\rho_w(z,t)\hat{\sigma}(z;t,0) = \bar{\sigma}(t)$. Substituting into (19) the values $s_\beta$ from (21) and $s_\gamma$ from (18), we obtain

$$\gamma \bar{T}_\gamma + \ln Z(\beta,\gamma) - \ln Z_\beta + \beta\bar{u} - \beta u_\beta = -<t-t_0>\bar{\sigma}_{st}. \tag{20}$$

Expressions (20) and (22) coincide at $\ln \rho_{rel}(u) = -\beta u - \ln Z_\beta$. In the quantities $\bar{u}$ and $u_\beta$, the random variable $u$ is averaged over different distributions: $\bar{u}$ over the distribution (11), and $u_\beta$ over the distribution $\rho_{rel}(u)$, $u_\beta = \bar{u}(\beta, \gamma = 0)$; the distribution density $\rho_{rel}(u)$ is equal to the function (11) at $\gamma=0$.

The definition of entropy (18) is generalized using expression (17), we assume

$$-\langle \ln \rho_{rel}(u) \rangle = \beta u_\beta + \ln Z_\beta = s_\beta, \quad Z_\beta = Z(\beta,\gamma)|_{\gamma=0}, \quad u_\beta = -\partial \ln Z_\beta / \partial \beta, \quad s_G = s_\beta; \quad (21)$$

$$s = \gamma \bar{T}_\gamma + \ln Z(\beta,\gamma) - <\ln \rho_{rel}> - \ln Z_\beta, \quad <\ln \rho_{rel}(u)> = <\ln \rho_{rel}(u)>_\rho = -\beta \bar{u} - \ln Z_\beta. \quad (22)$$

The averaging of $<\ln \rho_{rel}>$ in the ratio (22) is over the distribution (11), in the ratio (21) is over the distribution $\rho_{rel}(u)$.

Relation (10) is not fulfilled on the entire interval $t$-$t_0$. From the moment $t_0$ to some moment $t_1$ there is no stationarity, it comes after the moment $t_1$. We set $t_1$-$t_0$<<$t$-$t_1$. It is possible to take into account non-stationarity by expanding the expression $d\ln\rho_{rel}(t-y,-y)/dy$ into a series in $y=t-t_0$.

Wherein

$$Tr\{\rho_w \ln \rho_w(t)\} = s_G, \quad s = \bar{s}(t) = Tr\{\hat{s}(t,0)\rho_{rel}(t)\} = -\langle \ln \rho_{rel} \rangle, \quad \Delta s = s_G - s(t), \quad s_G = s_\beta. \quad (23)$$

The term "entropy" in the nonequilibrium case is largely arbitrary. This question is discussed in [77]. Expression (10) is rewritten as

$$-\Delta s = <\ln \rho_w(t)> - <\ln \rho_{rel}(t,0)>, \quad (24)$$

and expression (20) at $\bar{u}(\beta,\gamma) = u_\beta$ takes the form

$$-\Delta s = \gamma \bar{T}_\gamma + \ln Z(\beta,\gamma) - \ln Z_\beta = \gamma \bar{T}_\gamma + \ln Z_\gamma. \quad (25)$$

The value $\bar{u}(\beta,\gamma)$ from (15) is equal $u_\beta$ to (21) only in the case of independent variables $u, T_\gamma$, when the distribution density $f(T_\gamma)$ does not depend on $u$. In general, $f(T_\gamma, u)$. The average internal energy $\bar{u}(\beta,\gamma)$ can be divided into equilibrium $u_\beta$ and non-equilibrium $u_\gamma$ parts; $f(T_\gamma, u)$ is the probability density of the *FPT* distribution:

$$\bar{u} = -\frac{\partial \ln Z}{\partial \beta} = u_\beta + u_\gamma, \quad u_\beta = -\frac{\partial \ln Z_\beta}{\partial \beta}\bigg|_\gamma, \quad u_\gamma = -\frac{\partial \ln Z_\gamma}{\partial \beta}\bigg|_\gamma = \int_0^\infty e^{-\gamma T_\gamma}\left(\frac{\partial f(T_\gamma)}{\partial \beta}\right)dT_\gamma \frac{1}{Z_\gamma}.$$

For the case of independent variables $u, T_\gamma$, when $\bar{u}(\beta,\gamma) = u_\beta$, we write the term on the right side of (25) as

$$\ln Z(\beta,\gamma) - \ln Z_\beta = \ln \int e^{-\beta u}\omega(u)f(T_\gamma)e^{-\gamma T_\gamma}dudT_\gamma - \ln \int e^{-\beta u}\omega(u)du =$$
$$= \ln \int e^{-\beta u}\omega(u)f(T_\gamma)[1 - \gamma T_\gamma + \frac{1}{2}\gamma^2 T_\gamma^2 - ...]dudT_\gamma - \ln Z_\beta \quad (26)$$

Further transformations of relation (26) are affected by the fact whether the distribution $f(T_\gamma)$ depends on the random value of energy $u$. In the first approximation, if such a dependence is present, fluctuations of the random variable $u$ can be neglected by setting $u=\bar{u}(\beta,\gamma)$. Then the variables $u$ and $T_\gamma$ in (14), (16) are separated, and $Z(\beta,\gamma) = Z_\beta Z_\gamma$, $Z_\gamma = \int e^{-\gamma T_\gamma} f(T_\gamma)dT_\gamma$ the Laplace transform of the distribution $f(T_\gamma)$. In this case, expression (26) takes the form

$$\ln Z(\beta,\gamma) - \ln Z_\beta \approx -\gamma \bar{T}_0 + \frac{1}{2}\gamma^2 D_{T_0}, \quad D_{T_0} = \langle T_0^2 \rangle - \langle T_0 \rangle^2, \quad \bar{T}_0 = \bar{T}_0(\beta) = \bar{T}_{\gamma=0} = \bar{T}_\gamma(\beta,\gamma=0),$$

and relation (25) is written as

$$\Delta s \simeq \frac{1}{2}\gamma^2 D_{T_0}, \qquad \gamma = (\frac{2\Delta s}{D_{T_0}})^{1/2}. \tag{27}$$

Non-stationarity and time dependence from the past in expression (10) can be taken into account by expanding the function $\hat{\sigma}(t-y,-y)$ in powers $y=t-t_0$,

$$\hat{\sigma}(t-y,-y) = \hat{\sigma}(t,0) + ym_1 + ..., \quad m_1 = \frac{\partial \hat{\sigma}(t-y,-y)}{\partial y}\bigg|_{y=0}. \tag{28}$$

Then the left side of expression (27) after averaging takes the form

$$\Delta s = <t-t_0>\bar{\sigma}_{st}(1+<t-t_0>\bar{m}_1/\bar{\sigma}_{st}+...), \quad \bar{m}_1/\hat{\sigma}_{st} = \partial \ln \bar{\sigma}(t-y,-y)/\partial y\big|_{y=0},$$

and expression (27) is written as (in the general case, nonlinear dependences are valid)

$$\gamma = [\frac{2<t-t_0>\bar{\sigma}_{st}(1+<t-t_0>\bar{m}_1/\bar{\sigma}_{st}+...)}{D_{T_0}}]^{1/2}. \tag{29}$$

When specifying the value of $\Delta s$ in the form

$$\Delta s = \int_0^{\bar{T}_\gamma} \bar{\sigma}(t-y,-y)dy \simeq \bar{\sigma}(t)\bar{T}_\gamma + \frac{1}{2}(\bar{T}_\gamma)^2 \frac{\partial \bar{\sigma}(t-y,-y)}{\partial y}\bigg|_{y=0} + ... \approx \bar{\sigma}(t)\bar{T}_0 + \frac{1}{2}(\bar{T}_0)^2 \frac{\partial \bar{\sigma}(t-y,-y)}{\partial y}\bigg|_{y=0} + ...,$$

we obtain the following expression for $\gamma$:

$$\gamma = [\frac{2(\bar{\sigma}(t)\bar{T}_0 + \frac{1}{2}(\bar{T}_0)^2 \frac{\partial \bar{\sigma}(t-y,-y)}{\partial y}\big|_{y=0} + ...)}{D_{T_0}}]^{1/2}.$$

If we take into account cubic terms in $\gamma$ in expansion (26), then we obtain a cubic equation for $\gamma$ of the form

$$\Delta s = \frac{1}{2}\gamma^2 D_{T_0} - \frac{1}{3}\gamma^3(\langle T_0^3 \rangle + 2\langle T_0 \rangle^3 - 3\langle T_0 \rangle\langle T_0^2 \rangle). \tag{30}$$

When the value $\gamma$ in equation (30) is specified in the form $\gamma=\gamma_0+\varepsilon$, where $\gamma_0$ is determined by relation (27), and the value $\varepsilon$ is assumed to be a small addition to $\gamma_0$, we obtain a cubic equation for $\varepsilon$, the solution of which is sought in a linear approximation. The result for $\gamma$ is written as

$$\gamma \simeq \gamma_0[1-\gamma_0 \frac{1}{3}\frac{\partial^3 \ln Z_\gamma/\partial\gamma^3\big|_{\gamma=0}}{(\partial^2 \ln Z_\gamma/\partial\gamma^2\big|_{\gamma=0} + \gamma_0 \partial^3 \ln Z_\gamma/\partial\gamma^3\big|_{\gamma=0})}], \quad \gamma_0 = (\frac{2\Delta s}{D_{T_0}})^{1/2}.$$

The situation when the distribution $f(T_\gamma)$ depends on $u$ is illustrated by an example of a linear dependence, when

$$f(T_\gamma, u) = f_0(T_\gamma) + uf_1(T_\gamma), \quad f_1(T_\gamma) = \partial f(T_\gamma, u)/\partial u\big|_{u=0}. \tag{31}$$

Then

$$Z(\beta,\gamma) = \int e^{-\beta u}\omega(u)[f_0(T_\gamma) + uf_1(T_\gamma)][1 - \gamma T_\gamma + \frac{1}{2}\gamma^2 T_\gamma^2 - \frac{1}{6}\gamma^3 T_\gamma^3 + ...]dudT_\gamma.$$

Relations (27)-(30) will include $u_\beta$, $\langle u^2{}_\beta \rangle$, $\langle u^3{}_\beta \rangle$.

## 3. Comparison of EIT and TFPT

### 3.1. General comparison of EIT and TFPT

Compare *TFPT* results with *EIT*. In the *EIT* the independent thermodynamic variables are the energy $u_\beta$ and flows $q$ [8-13]. For simplicity, we restrict ourselves to the case of the simple problem of heat conduction in rigid bodies, without considering other transport processes. In *TFPT*, the variables are mean energy <u> and *FPT* $\bar{T}_\gamma$. The *EIT* [8-13] contains expressions out of (local) equilibrium up to terms of the second order in $\vec{q}$. In this approximation, an explicit expression for the entropy is also written:

$$s(u_\beta, q) = s_{eq}(u_\beta) - \frac{1}{2}\frac{\tau}{\rho\lambda\theta^2}\vec{q}\cdot\vec{q}, \tag{32}$$

$$ds = \theta^{-1}du_\beta - \frac{\tau}{\rho\lambda\theta^2}\vec{q}\cdot d\vec{q}, \qquad \theta^{-1} = T^{-1} - \frac{1}{2}\frac{\partial}{\partial u_\beta}(\frac{\tau}{\rho\lambda\theta^2})\vec{q}\cdot\vec{q}, \tag{33}$$

$$\sigma^s = \frac{1}{\lambda\theta^2}\vec{q}\cdot\vec{q}, \qquad a_\beta = \frac{\tau}{\rho\lambda\theta^2}, \tag{34}$$

$$\bar{u}_\beta = u_\beta = -\frac{\partial lnZ_\beta}{\partial\beta}\bigg|_\gamma, \quad Z_\beta = \int e^{-\beta u}\omega(u)du, \quad \bar{u} = \bar{u}(\beta,\gamma) = -\frac{\partial lnZ(\beta,\gamma)}{\partial\beta}\bigg|_\gamma, \tag{35}$$

where $\rho$ is the total mass density, $\lambda$ is the heat conductivity, $\theta$ is the non-equilibrium temperature [8], $q$ is the heat flux, $T$ is the local equilibrium temperature, $T(u)$, $u$ is the specific internal energy, $s$ is a local specific entropy, $\sigma^s$ is the entropy production rate, function $Z(\beta,\gamma)$ is defined in (14), (15), $Z_\beta = Z(\beta) = Z(\beta,\gamma)|_{\gamma=0}$, $s_{eq}(u_\beta) = s_\beta$ from (21).

Time $\tau$ from (32) is the relaxation time of heat fluxes. For very small values of the relaxation time $\tau$, the Maxwell-Cattaneo equation [8] (72) reduces to the Fourier law. In solid homogeneous bodies, during the time $\tau$, phonons or electrons transfer energy at the molecular level. This time interval $\tau$ is very small, it is approximately equal to the time between two successive collisions at the microscopic level. However, in polymers, superfluids, porous media, or organic tissues, slow internal degrees of freedom operate. In these cases, the time $\tau$ corresponds to the time required for the transfer of energy between different degrees of freedom. This time can be relatively large, on the order of or greater than 1 s [11]. The value of $\Delta s$ (19), (23) in *EIT* (32) is proportional to the relaxation time $\tau$. In this case, the process time is equal to $\tau$.

Consider expressions (18), (33) and equate them

$$ds_{EIT\tau} = \theta^{-1}d\bar{u}_\beta - \frac{\tau}{\rho\lambda\theta^2}\vec{q}\cdot d\vec{q} = ds_{TFPT} = \beta d\bar{u} + \gamma d\bar{T}_\gamma, \tag{36}$$

$$-d\Delta s = \beta(d\bar{u} - du_\beta) + \gamma d\bar{T}_\gamma, \tag{37}$$

where $\Delta s = \Delta s_{EIT\tau}$ is chosen in the form (32)-(34). If the variables $u$ and $T_\gamma$ are independent, then in (37), $d\bar{u} = du_\beta$, and

$$-d\Delta s_\tau = \gamma d\bar{T}_\gamma. \tag{38}$$

Let us substitute the expressions (18), (34) into the left side of (38),

$$\Delta = \Delta s = s_\beta - s = -[\gamma\bar{T}_\gamma + lnZ(\beta,\gamma) - \ln Z_\beta + \beta\bar{u} - \beta u_\beta] \to -(\gamma\bar{T}_\gamma + lnZ_\gamma),$$

where the arrow indicates the transition from the general case to the case of independent variables $u$ and $T_\gamma$. Differentiating in (38) with respect to $\gamma$ and to $\beta$, we obtain an identity.

Equality $\Delta s_{EIT\tau} = \Delta s_{TFPT}$, if in $a_\beta$ (32), (34) put $\theta = T$,

$$\Delta = \Delta s = s_\beta - s = \tau\vec{q}\cdot\vec{q}/2\rho\lambda T^2 = -[\gamma\bar{T}_\gamma + \ln Z(\beta,\gamma) - \ln Z_\beta + \beta\bar{u} - \beta u_\beta], \tag{39}$$

can be considered as an equation for expressing flows in terms of the parameter $\gamma$, $q(\gamma)$ or for finding the inverse relationship $\gamma(q)$.

## 3.2. Different probability models for FPT

From relations (13)-(16) it is clear that in order to determine the explicit form of the partition function and further determine thermodynamic quantities, it is necessary to know the *FPT* distribution. In [69], an exponential distribution was used, which can be compared to the hydrodynamic stage of the system's evolution [34-36]. It was noted in [70] that «Setting the form of the function плотности распределения *FPT* $p_\Gamma(y)$ reflects not only the internal properties of a system, but also the influence of the environment on an open system and the particular character of its interaction with the environment. The following physical interpretation of the exponential distribution for the function $p_\Gamma(y)$ is given: a system evolves freely like an isolated system governed by the Liouville operator. Besides that the system undergoes random transitions and the phase point representing the system switches from one trajectory to another one with an exponential probability under the influence of the "thermostat." The exponential distribution describes completely random systems. The influence of the environment on a system can have organized character as well; for example, this is the case of systems in a nonequilibrium state with input and output nonstationary fluxes. The character of the interaction with the environment can also vary; therefore different forms of the function $p_\Gamma(y)$ can be used». In [70] the maximum entropy principle for Liouville equations with source for the determination of the function $p_\Gamma(y)$ was applied. Obtaining explicit results using the distribution written in [70] faces mathematical difficulties. Therefore, in [70] series expansions are used, which are limited to several terms of the expansion, i.e., approximations. Various *FPT* distribution functions are written below to provide finite expressions for the partition function and thermodynamic functions.

Consider the case of exponential distribution for first-passage time; $T_0$ does not depend on $u$:
$$f(T_\gamma) = (T_0)^{-1} exp\{-T_\gamma/T_0\}. \tag{40}$$

The exponential distribution (40) is valid for large times, $t \to \infty$. In this case
$$Z_\gamma = 1/(1+\gamma T_0), \qquad \bar{T}_\gamma = \frac{T_0}{1+x}, \quad x = \gamma T_0, \qquad \frac{\partial \bar{T}_\gamma}{\partial \gamma} = -(\bar{T}_\gamma)^2, \tag{41}$$

$$D_{T_0} = (\bar{T}_\gamma)^2, \quad x = \gamma T_0 \approx \frac{\sqrt{a_\beta}\sqrt{\vec{q}\cdot\vec{q}}}{1-\sqrt{a_\beta}\sqrt{\vec{q}\cdot\vec{q}}} \approx \sqrt{a_\beta}\sqrt{\vec{q}\cdot\vec{q}}.$$

Substituting (27), (39) into (41) gives the relation between flows $q$ and *FPT*.
$$\bar{T}_\gamma = \frac{T_0}{1+\sqrt{\vec{q}\cdot\vec{q}}\sqrt{a_\beta}}, \quad \sqrt{\vec{q}\cdot\vec{q}} = \frac{1}{\sqrt{a_\beta}}(\frac{T_0}{\bar{T}_\gamma}-1). \tag{42}$$

For gamma distribution
$$f(x) = \frac{1}{\Gamma(\alpha)}\frac{1}{b^\alpha}x^{\alpha-1}e^{-x/b}, \quad x > 0, \quad f(x) = 0; \quad x < 0; \quad \int_0^\infty e^{-\gamma x}f(x)dx = (1+\gamma b)^{-\alpha}, \tag{43}$$

($\Gamma(\alpha)$ is gamma function) a special case of which for $\alpha=1$ is the exponential distribution (40),
$$Z_\gamma = 1/(1+\gamma T_0)^{-\alpha}, \quad \bar{T}_\gamma = \frac{T_0}{1+x/\alpha}, \quad x = \gamma T_0, \quad \bar{T}_\gamma = \frac{T_0}{1+\sqrt{\vec{q}\cdot\vec{q}}\sqrt{a_\beta/\alpha}}, \quad \sqrt{\vec{q}\cdot\vec{q}} = \frac{\sqrt{\alpha}}{\sqrt{a_\beta}}(\frac{T_0}{\bar{T}_\gamma}-1). \tag{44}$$

Entropy (18) for distribution (40), (41) is
$$s = s_\beta + \frac{x}{1+x} - \ln(1+x), \quad x = \gamma T_0. \tag{45}$$

Assuming the value of $s_{eq}$ from (32) equal to the value of $s_\beta$ (21), and comparing (32) and (45), we obtain the relation

$$\frac{x}{1+x} - \ln(1+x) = -a_\beta q^2/2 = -\Delta s_\tau, \quad a_\beta = \frac{\tau}{\rho\lambda\theta^2}. \tag{46}$$

Here $q^2$ stands for $\vec{q}\cdot\vec{q}$. If we put $\theta=T$ in $a_\beta$ (32), (34), then from (46) we obtain

$$\sqrt{\vec{q}\cdot\vec{q}} = [\frac{2\rho\lambda T^2}{\tau}(\ln(1+x) - \frac{x}{1+x})]^{1/2}, \tag{47}$$

expression for the flow $q$ in terms of $\gamma$ from $x=\gamma T_0$. For the parameter $\gamma$, relation (45) is a transcendental equation. Expanding the left-hand side of (46) into a series in powers of $x$, we obtain that for small $x$

$$x^2 = \gamma^2 T_0^2 = 2\Delta s. \tag{48}$$

For the exponential distribution (40) $D_{T_0} = T_0^2$, and expression (48) coincides with (27). Comparison of *TFPT* with the *NSO* method and with *EIT* leads to the same results. This is due to the fact that the expansion in powers of $\gamma$ of the expression for $s_\gamma = \ln Z_\gamma + \gamma\bar{T}_\gamma$ with independent parameters $u$ and $T_\gamma$, when $-s_\gamma = \Delta s = \gamma^2 D_{T_0}/2$, $D_{T_0} = \partial^2 \ln Z_\gamma / \partial\gamma^2|_{\gamma=0}$, is of a general nature.

Let us now consider the case of dependent variables $u$ and $T_\gamma$. For the exponential distribution (40), the parameter $T_0$ is the mean value. But averaging is carried out over the distribution of the random variable $T_\gamma$. No averaging is carried out over the parameter $u$. Therefore, in the general case, the dependence $T_0(u)$ is possible. It was assumed above that $T_0$ may depend on the average values of the energy $u$, and not on their random values. In cases of possible dependences of $T_0$ on $u$, random values of $u$ were replaced by their average values; fluctuations are neglected.

Consider the case of a linear dependence

$$T_0 = T_{00} + uT_{10}, \tag{49}$$

analogue (31), where $T_{00}$ and $T_{10}$ are some parameters. In this case, after integrating over $T_\gamma$ in (14), (16), (40), we obtain

$$Z(\beta,\gamma) = \int e^{-\beta u}\omega(u)\frac{du}{1+\gamma(T_{00}+uT_{10})}. \tag{50}$$

For small $\gamma$

$$Z(\beta,\gamma) \approx \int e^{-\beta u}\omega(u)[1-\gamma(T_{00}+uT_{10})]du = Z_\beta[1-\gamma(T_{00}+u_\beta T_{10})],$$

$$u_\beta = \frac{1}{Z_\beta}\int u e^{-\beta u}\omega(u)du, \quad Z_\beta = \int e^{-\beta u}\omega(u)du,$$

$$\ln Z(\beta,\gamma) \approx \ln Z_\beta + \ln[1-\gamma(T_{00}+u_\beta T_{10})], \tag{51}$$

$$\bar{u} = -\frac{\partial \ln Z(\beta,\gamma)}{\partial\beta}\bigg|_\gamma = u_\beta + \frac{\gamma(T_{10}\partial u_\beta/\partial\beta + u_\beta\partial T_{10}/\partial\beta + \partial T_{00}/\partial\beta)}{1-\gamma(T_{00}+u_\beta T_{10})}, \tag{52}$$

$$\bar{T}_\gamma = -\frac{\partial \ln Z(\beta,\gamma)}{\partial\gamma}\bigg|_\beta = \frac{T_{00}+u_\beta T_{10}}{1-\gamma(T_{00}+u_\beta T_{10})}, \quad T_0 = T_{00}+u_\beta T_{10}. \tag{53}$$

The average value of *FPT* is expressed in terms of flux:

$$\bar{T}_\gamma = \frac{T_0}{1-\gamma T_0} = \frac{T_0}{1-\sqrt{\vec{q}\cdot\vec{q}}\sqrt{a_\beta}}, \quad \sqrt{\vec{q}\cdot\vec{q}} = \frac{1}{\sqrt{a_\beta}}(1-\frac{T_0}{\bar{T}_\gamma}). \tag{54}$$

Suppose that in expression (50) the function $\omega(u) \approx \omega(\bar{u})$. Then for relation (50) we obtain

$$Z(\beta,\gamma) \approx -\omega(\bar{u})e^{\beta(1+\gamma T_{00})/\gamma T_{10}}\frac{1}{\gamma T_{10}}Ei(-\frac{\beta(1+\gamma T_{00})}{\gamma T_{10}}),$$

where $Ei(z)$ is an integral exponential function [78]. In the case of small values of $\gamma$, the function $E_1(z) = -Ei(-z)$ has an asymptotic expansion $E_1(z) \sim \frac{e^{-z}}{z}(1 - \frac{1}{z} + ...)$ [78]. Then $Z(\beta, \gamma) \simeq \omega(\bar{u}) \frac{1 - \gamma T_{10} / \beta(1 + \gamma T_{00})}{\beta(1 + \gamma T_{00})}$. Since at $\omega(u) \approx \omega(\bar{u})$, $Z_\beta = \omega(\bar{u})/\beta$, we can write some effective nonequilibrium reciprocal temperature of the form

$$\beta_{ef}(\beta, \gamma) \simeq \frac{\beta(1 + \gamma T_{00})}{1 - \gamma T_{10} / \beta(1 + \gamma T_{00})} \approx \beta(1 + \gamma T_{00}) + \gamma T_{10}, \quad \beta_{ef}(\beta, \gamma = 0) = \beta. \tag{55}$$

For dependence $T_0(u)$ of the form

$$T_0 = \frac{T_{00}}{1 + uT_{10}} \tag{56}$$

integration over $T_\gamma$ leads to the expression

$$\frac{1}{1 + \gamma T_0} = \frac{1}{1 + \gamma T_{00}/(1 + uT_{10})} \approx 1 - \gamma T_{00}/(1 + uT_{10}).$$

Then

$$Z(\beta, \gamma) = Z_\beta - \gamma T_{00} \int e^{-\beta u} \omega(u) \frac{du}{1 + uT_{10}}. \tag{57}$$

If the values of $T_{10}$ are small, then $Z(\beta, \gamma) \simeq Z_\beta[1 - \gamma T_{00}(1 - u_\beta T_{10})]$. Expressions for $\bar{u} = \bar{u}(\beta, \gamma)$ and $\bar{T}_\gamma = \bar{T}_\gamma(\beta, \gamma)$ are equal

$$\bar{u} = -\frac{\partial \ln Z(\beta, \gamma)}{\partial \beta}\bigg|_\gamma = u_\beta + \frac{\gamma((\partial T_{00}/\partial \beta)(1 - u_\beta T_{10}) - T_{00}(T_{10}\partial u_\beta / \partial \beta + u_\beta \partial T_{10}/\partial \beta))}{1 - \gamma T_{00}(1 - u_\beta T_{10})}, \tag{58}$$

$$\bar{T}_\gamma = -\frac{\partial \ln Z(\beta, \gamma)}{\partial \gamma}\bigg|_\beta = \frac{T_{00}(1 - u_\beta T_{10})}{1 - \gamma T_{00}(1 - u_\beta T_{10})}, \quad T_0 = T_{00}(1 - u_\beta T_{10}). \tag{59}$$

In the approximation $\omega(u) \approx \omega(\bar{u})$, the expression for $1/\beta$ in approximation (55) is replaced by

$$\frac{1}{\beta_{ef}(\beta, \gamma)} = \frac{1}{\beta} + \gamma e^{-\beta(1 + \gamma T_{00})/T_{10}} Ei(\frac{\beta(1 + \gamma T_{00})}{T_{10}}). \tag{60}$$

Let's consider another example. The exponential distribution (40) can be compared with the hydrodynamic stage of the evolution of a statistical system (in the terminology of [34–36]). For the preceding kinetic stage, an example of an inverse Gaussian or Wald distribution can be given (for example, [79], [45], [55]). In diffusion processes of the form (62) for *FPT*, the Wald distribution (for example, [45], [55]) $f(T_\gamma)$ has the form

$$F_{T_\gamma}(t) = \sqrt{\frac{|L|^2}{4\pi Dt^3}} \exp[-\frac{(L - vt)^2}{4Dt}], \quad \langle t \rangle = L/v, \quad a = v^2/D, \quad \bar{T}_0 = T_0 = \bar{T}_{\gamma=0} = \langle t \rangle. \tag{61}$$

where $L$ is the diffusion area in the one-dimensional case. Distribution (61) describes the drift-diffusion process. The Langevin equation for such a process has the form

$$\frac{dX(t)}{dt} = v + \zeta(t), \tag{62}$$

In equation (62), the term $\zeta(t)$ describes the Gaussian white noise. For it, the average value is zero, $\langle \zeta(t) \rangle = 0$ and the autocorrelation $\langle \zeta(t)\zeta(t') \rangle = D\delta(t - t')$ is related to the diffusion coefficient $D = k_B T/\varsigma$, $k_B$ is Boltzmann's constant, $T$ is the absolute temperature, $\varsigma$ is the friction coefficient.

The first term in (62) $v=F/\varsigma$ is the drift velocity, $F$ is an external force, for example, the drag force of the medium acting on a moving particle. The value $X(t)$ from (62) is a coordinate of the particle.

If we carry out the Laplace transform of distribution (61), then we obtain

$$Z_\gamma = \int_0^\infty e^{-\gamma t} F_{T_\gamma}(t) dt = \exp\{\frac{aT_0}{2}(1-\sqrt{1+4\gamma/a})\} \quad . \tag{63}$$

From relation (63) we obtain

$$\bar{T}_\gamma = -\frac{\partial \ln Z_\gamma}{\partial \gamma}\bigg|_\beta = \frac{T_0}{\sqrt{1+4\gamma/a}}, \quad T_0 = \langle t \rangle, \quad \bar{T}_\gamma = \frac{T_0}{\sqrt{1+(4\sqrt{\vec{q}\cdot\vec{q}}/a)\sqrt{aa_\beta/2T_0}}}, \quad \sqrt{\vec{q}\cdot\vec{q}} = \frac{a}{4\sqrt{aa_\beta/2T_0}}((\frac{T_0}{\bar{T}})^2-1) \tag{64}$$

(we took into account the expression for the dispersion obtained from (63)). *TFPT* uses dimensionless entropy, in units of $k_B$. From expression (39) with independent variables $u, T_\gamma$ when $\bar{u}(\beta,\gamma)=u_\beta$,

$$\Delta = \Delta s = \gamma \bar{T}_\gamma + \ln Z_\gamma = \frac{\gamma T_0}{\sqrt{1+4\gamma/a}} + \frac{aT_0}{2}(1-\sqrt{1+4\gamma/a}), \quad s_\beta = \beta u_\beta + \ln Z_\beta, \quad \bar{u} = -\frac{\partial \ln Z(\beta,\gamma)}{\partial \beta}\bigg|_\gamma \quad . \tag{65}$$

From (65) we find an expression for $\gamma(\Delta s)$ in the form

$$\Delta s = a_\beta q^2/2, \quad \gamma = \frac{2}{aT_0^2}p^2 - \frac{a}{2} \pm \frac{p}{T_0}\sqrt{\frac{4}{a^2}\frac{p^2}{T_0^2}-1}, \quad p = \Delta s + \frac{aT_0}{2}, \tag{66}$$

and the expression for $\Delta s(\gamma)$

$$\Delta s = a_\beta q^2/2 = \frac{aT_0}{2}(\frac{1+2\gamma/a}{\sqrt{1+4\gamma/a}}-1) \quad . \tag{67}$$

If the parameter $T_0$ in expressions (61), (63) depends linearly on the random variable $u$, as in (49), then from (49), (63)-(65) we obtain

$$\ln Z(\beta,\gamma) = \ln Z(\beta_{ef}) + (aT_{00}/2)(1-\sqrt{1+4\gamma/a}), \quad \beta_{ef} = \beta - (aT_{10}/2)(1-\sqrt{1+4\gamma/a}), \tag{68}$$

$$\bar{T}_\gamma = -\frac{\partial \ln Z(\beta,\gamma)}{\partial \gamma}\bigg|_\beta = \bar{u}(\beta_{ef})\frac{T_{10}}{\sqrt{1+4\gamma/a}} + \frac{T_{00}}{\sqrt{1+4\gamma/a}} = \frac{T_{0ef}}{\sqrt{1+4\gamma/a}}, \quad T_{0ef} = T_{00} + T_{10}\bar{u}(\beta_{ef}). \tag{69}$$

$$\bar{T}_\gamma = \frac{T_{0ef}}{\sqrt{1+(4\sqrt{\vec{q}\cdot\vec{q}}/a)\sqrt{aa_\beta/2T_{0ef}}}}, \quad \sqrt{\vec{q}\cdot\vec{q}} = \frac{a}{4\sqrt{aa_\beta/2T_{0ef}}}((\frac{T_{0ef}}{\bar{T}})^2-1), \tag{70}$$

$$\bar{u}(\beta_{ef}) = -\frac{\partial \ln Z(\beta_{ef})}{\partial \beta_{ef}}, \quad Z(\beta_{ef}) = Z_{\beta_{ef}} \quad .$$

Since $\beta_{ef}$ depends on $\gamma$, then (70) is an equation for $q$. We assume that the diffusion coefficient has the form $D = D_0 \exp[-\beta E_a]$, where $E_a$ is the activation energy. This value is included in the expression for $\bar{u}$, which is not written.

The relationship between $\gamma$ and $q$ takes on a more complex form than (65)

$$\Delta = \Delta s = \frac{1}{2}a_\beta q^2 = -[\gamma \bar{T}_\gamma + \ln Z(\beta,\gamma) - \ln Z_\beta + \beta \bar{u} - \beta u_\beta] = s_\beta - \frac{\gamma T_{0ef}}{\sqrt{1+4\gamma/a}} - \ln Z(\beta_{ef}) -$$

$$- \frac{aT_{00}}{2}(1-\sqrt{1+4\gamma/a}) + \beta \frac{\partial \ln Z(\beta,\gamma)}{\partial \beta}\bigg|_\gamma \quad . \tag{71}$$

Equating the values $\Delta s_{NSO}$, $\Delta s_{EIT}$, $\Delta s_{TFPT}$ has several aspects. In $\Delta s_{NSO}=<t-t_0><\sigma_{st}>$ (24) the main dependence on time $t$ is in the multiplier $<t-t_0>$. In $\Delta s_{EIT}$ (39), the fluxes depend on time. The Maxwell-Cattaneo equation [8] has the form

$$\vec{q} = -\lambda \nabla T - \tau \partial \vec{q}/\partial t. \qquad (72)$$

The solution of differential equation (72) is written as $\vec{q}(t) = -\int_0^t e^{-(t-t_1)/\tau} \frac{\lambda}{\tau} \nabla T(t_1) dt_1$, the dependence $\lambda(t)$, $\tau(t)$ is possible. Entropy change (32) is $\Delta s_{EIT} = \tau \Delta_1$, $\Delta_1 = \vec{q} \cdot \vec{q}/2\rho\lambda\theta^2$.

Perhaps one should interpret the relaxation time parameter $\tau$ of the Maxwell-Cattaneo equation (72) as the average $FPT$ $T_\gamma$. In this case, the value $\tau=T_\gamma$ is not a constant value, but depends on the parameter $\gamma$, and through the parameter $\gamma$ and on the change in entropy $\Delta s$ and flows $\vec{q}$.

The possibility of considering the variables $u$ and $T_\gamma$ independent remains an open question. The independence of the distribution of $FPT$ $T_\gamma$ on the random variable $u$ [45] seems rather artificial. This is an approximation connected with the replacement of the random variable $u$ by the mean value $<u>$. It is also necessary to investigate the question of what physical conditions correspond to different models of the dependences of the $FPT$ distribution on the random value of energy $u$.

Another important question is about the time constant $\tau$ in (32)-(34). Replacing the time constant $\tau$ by $\bar{T}_\gamma$ in $EIT$ makes it possible to obtain meaningful thermodynamic relations relating the fluxes from $EIT$ with the parameter $\gamma$ from $TFPT$ conjugate to $T_\gamma$. And it's physically justified. The process time is not constant, but itself depends on the process. In the proposed approach, the mean $FPT$ $\bar{T}_\gamma$ depends on $\Delta s$, the change in entropy, which in turn depends on $\bar{T}_\gamma$. The resulting expressions are self-consistent: the change in entropy depends on the average value of $FPT$, which depends on the change in entropy. What thresholds are reached by $FPT$? In [80] it is written: «The observable $O$, the threshold $D$, and the time $\tau_F := \inf\{t \geq 0 : O(\omega t) \in D\}$, define the physical process and its duration. For concreteness, $\tau_F$ may be the minimum time to displace a mass, or to exchange a given amount of energy or particles with a reservoir. In general, it represents the time needed for a specific physical process to be carried out by the system». This quantity $\tau_F$ differs significantly from the relaxation time $\tau$. The deviation of the stationary value of fluxes that exceeds the standard deviation, as well as a number of other physical quantities, for example, the time to reach a new stationary or equilibrium state, can be considered as a threshold.

In [80], $FPT$ is used in entropy production $\Delta s = \bar{\tau}_F \langle ds/dt \rangle$. We changed the notation $\tau$ in [80] to $\tau_F$ to avoid confusion with $\tau$ from (32)-(24), (72).

The parameter $\tau$ in (32) depends on $\beta$. As in [80], we will consider this value as mean $FPT$, setting

$$\tau \rightarrow \bar{\tau}(\beta,\gamma) = \bar{T}_\gamma(\beta,\gamma). \qquad (73)$$

Assumption (73) combines the time parameters of $EIT$ and $TFPT$.

Let us show by examples that $\bar{\tau}(\beta,\gamma)$ in $\Delta s_{EIT}$ it depends only on $\beta$. For the exponential distribution (40) and relation (46), we take $\Delta s = \Delta s_{EIT} = \bar{T}_\gamma \Delta_1$, rather than $\tau \Delta_1$, and write

$$\Delta s = \Delta s_{EIT} = \bar{T}_\gamma \Delta_1 = \ln(1+x) - x/(1+x), \quad x = \gamma T_0, \quad \Delta_1 = \vec{q} \cdot \vec{q}/2\rho\lambda\theta^2.$$

Expanding this expression into a series, we obtain relation (27) with $\Delta s = T_0 \Delta_1$, the value of $\tau$ is replaced by $T_0(\beta)$, which does not depend on $\gamma$. We obtain the same for the Wald distribution (61) and relation (65) with $\tau$ replaced by $T_0(\beta)$. If we expand in general both parts of relation (39), where $\Delta s = \bar{\tau}_\gamma(\beta,\gamma)\Delta_1 = \Delta_1[T_0 + \gamma D_{T_0} + \gamma^2(\partial D_{T_\gamma}/\partial\gamma_{|\gamma=0})/2+...]$, then, up to $\gamma^2$, we obtain a quadratic equation for $\gamma$ with the solution

$$\gamma = \frac{1}{(D_{T_0} - \Delta_1(\partial D_{T_\gamma}/\partial\gamma_{|\gamma=0}))}[\Delta_1 D_{T_0} \pm \sqrt{(\Delta_1 D_{T_0})^2 + 2\Delta_1 T_0(D_{T_0} - \Delta_1(\partial D_{T_\gamma}/\partial\gamma_{|\gamma=0}))}]. \quad (74)$$

Assuming in (74) the small parameter entropy production $\Delta_1$, we obtain for $\gamma$ the expression

$$\gamma = \frac{\sqrt{2\Delta_1 T_0 D_{T_0}}}{D_{T_0}}, \quad \gamma^2 = \frac{2\Delta_1 T_0}{D_{T_0}}, \quad (75)$$

that is, expressions (27), (48) with $\Delta s = T_0 \Delta_1$.

## 4. Physical meaning of the parameter γ. Other expressions for γ

Above, the parameter $\gamma$ was associated with a change in the internal entropy of the system. In the general case, changes in entropy caused by the exchange of entropy with the environment should be taken into account, as was done, for example, in [81]. In addition to the above expression (27), (48) for the parameter $\gamma$ through a change in entropy, there are other possibilities for determining the parameter $\gamma$. Let's point out some of them.

1. From (18) we obtain

$$\gamma = \frac{\partial s_\gamma}{\partial\langle T_\gamma\rangle}\bigg|_{\langle u\rangle} = -\frac{\partial\langle u\rangle}{\partial\langle T_\gamma\rangle}\bigg|_{s_\gamma}\left(\frac{\partial\langle u\rangle}{\partial s_\gamma}\right)^{-1}_{\langle T_\gamma\rangle} = -\frac{1}{T}\frac{\partial\langle u\rangle}{\partial\langle T_\gamma\rangle}\bigg|_{s_\gamma}, \quad \frac{\partial\langle u\rangle}{\partial\langle T_\gamma\rangle}\bigg|_{s_\gamma} = \frac{\partial\langle u\rangle}{\partial\beta}\bigg|_\gamma \frac{\partial\beta}{\partial\langle T_\gamma\rangle}\bigg|_{s_\gamma} + \frac{\partial\langle u\rangle}{\partial\gamma}\bigg|_\beta \frac{\partial\gamma}{\partial\langle T_\gamma\rangle}\bigg|_{s_\gamma}. \quad (76)$$

It is possible to further transform expression (76).

2. Taking into account the dependence of the *FPT* distribution density on random energy cancels the simple picture of independent variables and opens up possibilities for a more in-depth description. The article considers simple cases of linear dependence on random energy. In this case, the description becomes more complicated, new dependences on the parameter γ appear, and through it on changes in entropy and fluxes. The type of explicit dependence of the *FPT* distribution on random energy depends on the specific problem. Let us use the thermodynamic uncertainty relation (*TUR*) for *FPT*, a quantity associated with flow, use *TUR* [82, 83, 84] according to which $2T_0^2/D_{T0} \sim T_0\bar{\sigma}$, then substituting this expression into expression (27) with $\Delta s = \sigma T_0$, we obtain

$$\gamma \sim \bar{\sigma}. \quad (77)$$

The sign of proportionality, and not equality, is here because, firstly, expression (27) is approximate, and, secondly, *TUR* gives only estimated relationships.

Let us now consider another estimate for the value of γ. Approximation (10) for *NSO* is valid for the stationary case, when the entropy production operator does not depend on time, and for the case when $\hat{\sigma}(t-y,-y) \simeq \hat{\sigma}(t)$, a function that does not depend on the past $y=t-t_0$. Then in the second term on the right side of expression (1), the entropy production operator is taken out of the integration, and expression (1) takes the form $ln\rho_w(t)=ln\rho_{rel}(t,0)-<t-t_0>\hat{\sigma}_{st}$ (10). In this expression, let us compare the distribution $\rho_w(t)$ with the distribution (11)-(14), and the distribution $\rho_{rel}(t,0)$ with the distribution with $\rho_{rel} = \exp[-\beta u]/Z_\beta, \; Z_\beta = \int\exp[-\beta u]dz$. Then in (10)

$\langle t-t_0 \rangle \bar{\sigma}_{st} = \ln Z(\beta, \gamma/Z_\beta) + \gamma \langle T_\gamma \rangle$. From here $\langle t-t_0 \rangle \bar{\sigma}_{st} = \ln Z((\beta, \gamma/Z_\beta)) + \gamma \langle T_\gamma \rangle$, $\gamma = \langle t-t_0 \rangle \bar{\sigma}_{st} / \langle T_\gamma \rangle + \ln Z(\beta, \gamma/Z_{\beta/}))/\langle T_\gamma \rangle$. At large $\langle T_\gamma \rangle$ and $\langle t-t_0 \rangle \sim \langle T_\gamma \rangle$, $\gamma \sim \bar{\sigma}_{st}$.

Another case of expressing the function $\gamma$ from relation (27) $\gamma = (2\bar{T}/D_{T_0})^{1/2} \sigma^{1/2} \sim \sigma$ can be obtained using the relations of extended nonequilibrium thermodynamics (*EIT*), namely, the expression $\sigma^s = \frac{1}{\lambda \theta^2} \vec{q} \cdot \vec{q}$ (34) and *TUR* $2T_0^2/D_{T0} \sim T_0 \bar{\sigma}$. Substituting this expression and $\sigma = \sigma^S$ into (27), we obtain $\gamma = (2\Delta s / D_{T_0})^{1/2} = (2\sigma^2)^{1/2} \sim \sigma$. From (27) and the expression $\sigma^s = \frac{1}{\lambda \theta^2} \vec{q} \cdot \vec{q}$ for $\theta \approx T$ we obtain that

$$\gamma = (2T_0 / D_{T_0} \lambda T^2)^{1/2} \sqrt{\vec{q} \cdot \vec{q}} = a\sqrt{\vec{q} \cdot \vec{q}}, \quad a = (2T_0 / D_{T_0} \lambda T^2)^{1/2}, \tag{78}$$

where $\lambda$ is the thermal conductivity coefficient, $T$ is the temperature. The above recording is not strict, but evaluative, semi-qualitative.

From a comparison with *NSO* we obtain the proportionality of the parameter $\gamma$ to the entropy production $\sigma$ (77), and from a comparison with *EIT* we obtain the proportionality of the parameter $\gamma$ to the flows $q$ (78). In the case of *NSO* and expression (77), small values of the parameter $\gamma$ and large *FPT* times are estimated, and memory effects are not taken into account. For *EIT*, expression (78) is also valid for small values of the parameter $\gamma$; an expression of the form $\sigma = \sigma^S$ (34) from *EIT* is considered. Apparently, it is necessary to conduct a more detailed and detailed mathematical study of these cases. Here we carry out a qualitative estimate of the parameter $\gamma$. The importance of the parameter $\gamma$, comparable to the temperature parameter, was noted above. Unfortunately, this parameter $\gamma$ is practically unknown and does not appear in any studies. The physical situation with the parameter $\gamma$ may be related to what time scale the system is on, what physical situation is being considered, etc.

3. Another possibility for estimating the parameter $\gamma$ is related to thermodynamics of trajectories [85-90, 94-99]. An ensemble of dynamic trajectories is considered, similar to the thermodynamic method ensembles for configurations or microstates of equilibrium statistical mechanics. Order parameters are observable in time, the fluctuation behavior of which characterizes the dynamics of the system. Large size or the thermodynamic limit in the case of dynamics also includes the limit of a long observation time. This is the large deviation mode [91], in which the large deviation (*LD*) functions play a role dynamic entropy or free energy [91, 88]. We consider *s*-ensembles in which the process time is fixed, and, for example, the dynamic activity $K$ and the number of trajectory changes fluctuate. In this case the dynamical analog of thermodynamic partition function is

$$Z_\tau(s) \equiv \sum_K e^{-sK} P_\tau(K) = \sum_{X_\tau} e^{-s\hat{K}[X_\tau]} P(X_\tau), \quad Z_\tau(s) \sim e^{\tau\theta(s)}, \tag{79}$$

where a trajectory is denoted by $X_\tau$, $K$ is dynamic activity defined as the total number of configuration changes per trajectory, the probability $P_\tau(K)$ is distribution of $K$ over all trajectories $X_\tau$ of total time $\tau$, the probability $P(X_\tau)$ is the probability to observe this trajectory out of all the possible ones of total time $\tau$, the function $\theta(s)$ is considered as a (negative) dynamic free energy per unit time, the field s conjugate to $K$. For large $\tau$ the generating function also acquires a *LD* form. In [89], *x*-ensembles are introduced in which $K$ is fixed and the time fluctuates. The corresponding moment generating function for $\tau$ is

$$Z_K(x) \equiv \int_0^\infty d\tau e^{-x\tau} P_K(\tau) = \sum_{Y_K} e^{-x\hat{t}[Y_K]} P(Y_K), \tag{80}$$

where $P(Y_K)$ is the probability of a trajectories $Y_K$, $Y_K = (C_0 \to C_{t_1} \to ... \to C_K)$, where $C_0$ is the initial configuration and $t_i$ the time when the transition from $C_{t_{i-1}}$ to $C_{t_i}$ occurs (so that the waiting time for the $i$-th jump is $t_i - t_{i-1}$), $P_K(\tau)$ is equal to the distribution of the total trajectory length $\tau$ for a fixed activity $K$, $\langle \tau^n \rangle = (-1)^n \partial_x^n Z_K(x)|_{x=0}$. For large $K$ the generating function also has a LD form [91],

$$Z_K(x) \sim e^{Kg(x)}. \tag{81}$$

Equation (80) is the partition sum for the ensemble of trajectories with probabilities
$$P_x(Y_K) \equiv Z_K^{-1}(x) e^{-x\hat{t}[Y_K]} P(Y_K). \tag{82}$$
The function $g$ is the functional inverse of $\theta$ and vice versa [89]
$$\theta(s) = g^{-1}(s), \quad g(\gamma) = \theta^{-1}(\gamma), \quad s = g(\gamma), \quad \gamma = \theta(s). \tag{83}$$
We replace $g(x)$ with $g(\gamma)$ (to use the notation $\gamma$ from [44-48] for the parameter conjugate to $FPT$). Relations (83) give expressions for $\gamma$.

In [90] consider trajectory observables defined in terms of the jumps in a trajectory
$$\theta(s) = \bar{k}(e^{-s\langle a \rangle / \bar{k}} - 1), \quad \bar{k} = \sum_{xy} \rho_x w_{xy}, \quad \langle a \rangle = \sum_{xy} \alpha_{xy} \rho_x w_{xy}. \tag{84}$$

where $\rho_x$ is the stationary distribution. In [90] consider continuous-time Markov chain $X := (X_t)_{t>0}$ taking values in the finite state space $E$ with generator $W = \sum_{x \neq y} w_{xy} |x\rangle\langle y| - \sum_x w_{xx} |x\rangle\langle x|$, with $x, y \in E$. We assume that $X$ is irreducible with unique invariant measure (i.e., stationary state) $\rho$. We are interested in studying fluctuations of observables of the trajectory $X$ of the form
$$A(\omega) = \sum_{xy} \alpha_{xy} Q_{xy}(\omega), \tag{85}$$

where $Q_{xy}(\omega)$ is the number of jumps from $x$ to $y$ in trajectory $\omega$. We will assume all $\alpha_{xy} \geq 0$. This means that $A(\omega)$ is non-negative and nondecreasing with time.

From (84) we have
$$\gamma = \theta(s) = \bar{k}(e^{-s\langle a \rangle_{\pi+} / \bar{k}} - 1) \approx \bar{k}(-s\langle a \rangle_{\pi+} / \bar{k} + ...).$$

In this case, the parameter $\gamma$ is expressed not through the change in entropy, but through the parameter $s$. If we use equality (79), $\theta(s) = \ln Z_\tau(s)/\tau$, and assume that the parameter $\tau$ is equal to the average value depending on $\gamma$, use the expression for the average $\tau$ obtained in [81] for the $\theta(s)$ form (84), $\langle \tau_\gamma(x) \rangle = \dfrac{x}{\langle a \rangle_{\pi+}} \dfrac{1}{(1+\gamma/\langle k \rangle)}$, $x$ is the value that is achieved by the random process (85), then we obtain an equation for $\gamma$ with the solution
$$\gamma = \frac{(s_\tau - s\bar{K})/\tau_0}{(1-(s_\tau - s\bar{K})/\tau_0 \bar{k})}, \quad \tau_0 = x/\bar{a}. \tag{86}$$

In expression (86) $x = \langle K \rangle$, and the quantity $\ln Z_\tau(s)$ is expressed in terms of free energy and in terms of $s_\tau - s\langle K \rangle$, where $s_\tau$ is the entropy of the distribution with thermodynamic partition function (79). Thus, the parameter $\gamma$ is also associated with entropy, but a special case of it. For other functions $\theta(s)$, for example, for a two-level classical system [89, 90, 92], similar relations are valid. The entropy $s_\tau$ remains unknown. Calculating $s_\tau$ as Shannon entropy leads again to $\ln Z_\tau(s)$. This value should be assessed from other considerations. The situation is similar for determining the parameter $\gamma$ from expression (81), where the partition function $Z_K(\gamma)$ can be specified as the Laplace transform of the probability density of the *FPT* distribution. For example, for a two-level classical system [89, 90, 92] you can specify a superposition of two exponential distributions. Then from expression (81) we obtain an equation for determining the parameter $\gamma$. If you set the boundary *x=A*, where *A* is the observable from (85), then the parameter $\tau_0 = \langle \tau(\gamma) \rangle |_{\gamma=0}$ remains unknown, which will also need to be found from other conditions.

## 5. Conclusion

Any physical theory contains some ideas. In the *NSO* approach [34-44], this is the consideration of the history of the system and a rigorous description of kinetic processes and equations of hydrodynamics. Information statistical thermodynamics [18-33], developed on the basis of *NSO*, describes in detail the physical prerequisites underlying both *NSO* and information statistical thermodynamics. Its applications are very wide [27-33]. *EIT* contains non-local and non-linear effects, memory effects, as in *NSO*. In *TFPT*, the finiteness of the time of physical processes occurring in physical systems and the finiteness of the lifetime of the systems themselves are laid down; the importance of *FPT* is emphasized in *TFPT*. *FPT*s are related and over time. Any time is *FPT*, reaching some threshold. *FPT* finds wide applications in many physical, chemical, biological and other problems [49–52, 54–68]. *TFPT* also has applications [93-95]. For example, expression (27) was used in [82] in determining and estimating the period of a nuclear reactor.

The version of non-equilibrium thermodynamics containing *FPT*, *TFPT*, is consistent with the *EIT* and *NSO* method. The latter theories are true far from equilibrium and have considerable generality. Appendix A shows that, at a constant average *FPT*, *TFPT* has the same thermodynamic properties as the equilibrium theory. *TFPT* has a wide spectrum of possibilities: far from equilibrium, like *NSO*, and equilibrium properties. Theories of *EIT* and *NSO* method are applicable to almost all physical systems. However, *TFPT* is also applicable to equilibrium phenomena (Appendix A). Perhaps, due to the mathematical generality of the definition of *FPT* [72-74] used in *TFPT* (relations (11)-(14)), the theory of *TFPT*, based on *FPT*, can be applicable both in the equilibrium description and in the description of states that are far from equilibrium. If energy is the main thermodynamic parameter at equilibrium, then *FPT* can be the main thermodynamic parameter at non-equilibrium. The equations for the *FPT* distribution are conjugate to the equations for the energy distribution [96].

Many features of *EIT*, such as the centrality of entropy, additional thermodynamic parameter, are shared by *TFPT*. A quantity similar to *FPT* in *TFPT* is implicitly contained in the nonequilibrium statistical operator (*NSO*) method [34–44]; in it, the quasi-equilibrium distribution is averaged over the distribution of the lifetime of the system [44]. Representing the relaxation time $\tau$ in *EIT* as *FPT* brings all three theories closer together. The proposed version of non-equilibrium thermodynamics assumes that the Laplace transform function of the probability density distribution *FPT* is known, which limits the generality of the theory. However, in equilibrium thermodynamics, in order to apply it to specific systems, one must know the equation of state and other characteristics of the system. There are a lot of *FPT* distributions, as well as random processes themselves. Classes of this kind of distributions can be divided according to the stages of the evolution of the system [34-36]. In a specific problem, it is necessary to find a random process that best suits this problem. In many cases it is difficult to find an exact solution for the *FPT* distribution or its Laplace transform. Then it is possible to use the approximate approaches obtained in a number of papers. For example, in [66–67], expansions were obtained for the *FPT* distribution and its Laplace transform, the convergence of the series was estimated, and estimates were made of how much the next term of the series is less than the previous one. There are other possibilities for characterizing the *FPT*. The connection between the stage of evolution at which the system under consideration is located, using various approximations, is considered in [97, 84].

Two examples of distributions are considered: the exponential distribution and the Wald distribution. The exponential distribution was used in the nonequilibrium statistical operator [34-36]. In [80], the exponential distribution for *FPT* was also used. This was justified by the rarity of events. The exponential distribution can be compared with the hydrodynamic stage of the evolution of the system (in the terminology of [34-36]), and the Wald distribution (many other distributions can be used for this) with the kinetic stage.

By itself, the value of the average *FPT* $\bar{T}_\gamma$ is informative and important in many areas of science and its applications [49-52, 54-68]. And the expression of this quantity in terms of the entropy change parameters $\Delta s$, flows $q$ and $\gamma$ should be useful. The parameters $\Delta s$ and $q$ are known and important. The parameter $\gamma$ is also important. Its meaning is not entirely clear. Based on the symmetry of the thermodynamic parameters $u$ and $T_\gamma$ from (11), this parameter is similar to the parameter $\beta$ - the reciprocal temperature. The equation for the distribution of $T_\gamma$ is conjugate to the equation for the energy $u$ [96], the basic quantity of equilibrium thermodynamics. If $\gamma \sim \sqrt{\sigma^s} \sim q$ proportional to the root from entropy production and flow, then the physical meaning of this parameter is perhaps more transparent than that of $\beta$? A detailed mesoscopic description can be obtained using explicit stochastic models of the system, such as diffusion type [49-52] or stochastic storage processes [99]. The latter can be considered as models of a system excited by generalized noise. They take into account the interaction with the thermostat (reservoir), as well as the moments of degeneracy and loss of stationarity under certain conditions.

The introduction of additional state variables is a common feature of *EIT* and *TFPT*. At the same time, fluxes and *FPT* have a lot in common. Therefore, it is interesting to compare these areas of nonequilibrium thermodynamics. With *EIT*, the approach proposed in [69-70] is united by the fact that in both cases an additional thermodynamic parameter is introduced. In *EIT* it is flows, in thermodynamics with first-passage time (*TFPT*) it is *FPT*. A similar value is implicitly contained in the method of nonequilibrium statistical operator (*NSO*) [34-44]; in it, the quasi-equilibrium distribution is averaged over the distribution of the past lifetime of the system [44].

The relation (27) obtained, which relates the parameter $\gamma$ to the change in entropy and fluxes, is of great importance. Using (27), expressions (42), (44), (54), (64), (70) are written, expressing the average *FPT* in terms of flows, and inverse expressions: flows in terms of the average *FPT*. Like the *NSO*, the theory is self-consistent: the average *FPT* is expressed in terms of the change in entropy, and the change in entropy depends on the average *FPT*.

Distributions containing *FPT* as a thermodynamic parameter are widely used in the thermodynamics of trajectories (*x*-ensembles [89, 100]), [90]. In [81], the approaches used in [45-48] are applied to the thermodynamics of trajectories.

Dependences of the *FPT* distribution density on random energy change the simplest picture of independent variables and create opportunities for a more detailed and in-depth description. The article considers the simplest case of linear dependence. It illustrates how the description becomes more complicated, new dependences on the parameter $\gamma$ appear, and through it on the change in entropy and flows. Setting the dependence of the *FPT* distribution on random energy depends on the specific task.

## Appendix A. Coincidence of TFPT relations with equilibrium expressions at a constant value of the average FPT $<T_\gamma>$

When given in (18)
$$\beta F_{T\gamma}=-[lnZ(\beta,\gamma)+\gamma<T_\gamma>], \quad s=\beta(<u>-F_{T\gamma}); \quad ds=\beta d<u>+\gamma d<T_\gamma>, \quad (A1)$$
at a constant value of the average *FPT* $<T_\gamma>$ from (18), (A1) we obtain thermodynamic relations that coincide with the equilibrium:
$$\beta=1/T=\partial s/\partial<u>|_{<T_\gamma>}; \quad s=-\partial F_{T\gamma}/\partial T|_{<T_\gamma>}; \quad <u>=F_{T\gamma}-T\partial F_{T\gamma}/\partial T|_{<T_\gamma>}; \quad c_v=\partial<u>/\partial T|_{<T_\gamma>}=T\partial s/\partial T|_{<T_\gamma>},$$
where
$$c_v=\beta^2 D/\Delta_{T\gamma}; \quad D=\Delta_u\Delta_{T\gamma}-\Delta^2; \quad \Delta_u=-\partial<u>/\partial\beta|_\gamma=<u^2>-<u>^2; \quad (A2)$$
$$\Delta_{T\gamma}=-\partial<T_\gamma>/\partial\gamma|_\beta=<T_\gamma^2>-<T_\gamma>^2; \quad \Delta=-\partial<u>/\partial\gamma|_\beta=-\partial<T_\gamma>/\partial\beta|_\gamma=<T_\gamma u>-<T_\gamma><u>,$$
(averaging is carried out over the distribution (11)), $c_v$ in (A2) is an analog of the equilibrium heat capacity; it can also be $c_p$, since the volume coordinate is not considered.

For spatially inhomogeneous systems, the quantities $\beta$ and $\gamma$, generally speaking, depend on the spatial coordinate. In nonequilibrium thermodynamics, including *EIT*, the densities of extensive thermodynamic quantities such as entropy, internal energy, mass fraction of a component, etc. are considered. We adhere to this approach, including for *FPT*.

## References

1. T. D. Donder, *L'Affinitè* (Gauthier-Villars, Paris, France, 1936).
2. L. Onsager, Reciprocal Relations in Irreversible Processes. *Phys. Rev.* **37**, 405 (1931).
3. I. Prigogine, *Etude Thermodinamique des Phènomènes Irrèversibles* (Desoer, Liege, Belgium, 1947).
4. L. Onsager and S. Machlup, Fluctuations and Irreversible Processes. *Phys. Rev.* **9**, 1505 (1953).
5. S. de Groot and P. Mazur, *Nonequilibrium Thermodynamics* (North Holland, Amsterdam, The Netherlands, 1962).
6. P. Glansdorff and I. Prigogine, *Thermodynamic Theory of Structure, Stability, and Fluctuations* (Wiley-Interscience, New York, USA, 1971).
7. C. Truesdell, *Rational Thermodynamics* (McGraw-Hill, New York, USA, 1985), [second enlarged edition (Springer, Berlin, Germany, 1988)].
8. D. Jou, J. Casas-Vazquez and G. Lebon, *Extended Irreversible Thermodynamics*, 1993. First edition. Second edition, 1996. Third edition 2001. Fourth edition 2010. Springer, Berlin.
9. G. Lebon, D. Jou, J. Casas-Vázquez, *Understanding Non-equilibrium Thermodynamics, Foundations, Applications, Frontiers*, Springer, Berlin, Heidelberg, 2008.
10. J. Casas-Vázquez, D. Jou and G. Lebon. (eds.), *Recent Developments in Non-Equilibrium Thermodynamics*. In: Lecture Notes in Physics, Vol. 199. Springer, Berlin, 1984.
11. G. Lebon & D. Jou, Early history of extended irreversible thermodynamics (1953–1983): An exploration beyond local equilibrium and classical transport theory, *The European Physical Journal H*, **40**, 205–240 (2015).

12. B. C. Eu, *Kinetic Theory of Irreversible Thermodynamics* (Wiley, New York, USA, 1992). B. C. Eu, *Nonequilibrium Statistical Mechanics. Ensemble Method, Fundamental Theories of Physics*, 93, Kluwer, Dordrecht, 1998.
13. I. Muller and T. Ruggeri, *Extended Thermodynamics* (Springer, Berlin, Germany, 1993).
14. R. V. Velasco and L. S. García-Colín, The kinetic foundations of non-local nonequilibrium thermodynamics, *J. Non-Equilib. Thermodyn.*, **18**, 157 (1993).
15. I. Gyarmati, The wave approach of thermodynamics and some problems of non-linear theories. *J. Non-Equil. Thermodyn*. **2**, 233-260 (1977).
16. M. Grmela, Thermodynamics of driven systems, *Phys. Rev. E*, **48**:2, 919–930, (1993).
17. U. Seifert, Stochastic thermodynamics: principles and perspectives. *The European Physical Journal B*. **64** (3–4): 423–431 (2008). arXiv:0710.1187.
18. A. Hobson, Irreversibility and Information in Mechanical Systems. *J. Chem. Phys*. **45**, 1352 (1966).
19. L. S. Garcia-Colin, Á. R. Vasconcellos, and R. Luzzi, On Informational Statistical Thermodynamics, *J. Non-Equilib. Thermodyn*. **19**, 24 (1994).
20. R. Luzzi, Á. R. Vasconcellos, and J. G. Ramos, *Statistical Foundations of Irreversible Thermodynamics* (Teubner-BertelsmannSpringer, Sttutgart, Germany, 2000).
21. J. R. Madureira, Á. R. Vasconcellos, R. Luzzi, L. Lauck, Markovian kinetic equations in a nonequilibrium statistical ensemble formalism, *Phys. Rev. E*, **57**:3 (1998), 3637–3640.
22. J. R. Madureira, Á. R. Vasconcellos, R. Luzzi, J. Casas-Vazquez, D. Jou, Evolution of dissipative processes via a statistical thermodynamic approach. I. Generalized Mori–Heisenberg–Langevin equations, *J. Chem. Phys*., **108**:18, 7568–7579 (1998).
23. J. G. Ramos, Á. R. Vasconcellos, R. Luzzi, A classical approach in predictive statistical mechanics: a generalized Boltzmann formalism, *Fortschr. Phys*., **43**:4, 265–300 (1995).
24. F. S. Vannucchi, Á. R. Vasconcellos, R. Luzzi, Thermo-statistical theory of and relaxation processes, *Internat. J. Modern Phys*. B, **23**:27, 5283–5305 (2009).
25. Á. R. Vasconcellos, R. Luzzi, J. G. Ramos, Irreversible thermodynamics in a nonequilibrium statistical ensemble formalism, *La Rivista del Nuovo Cimento*, **24**:3, 1–70 (2001).
26. R. Luzzi, Á. R. Vasconcellos, J. G. Ramos, The theory of irreversible processes: foundations of a non-equilibrium statistical ensemble formalism, *La Rivista del Nuovo Cimento*, **29**:2, 1–82 (2006).
27. R. Luzzi, Á. R. Vasconcellos, J. G. Ramos, Non-equilibrium statistical mechanics of complex systems: an overview, *La Rivista del Nuovo Cimento*, **30**:3, 95–157 (2007).
28. C. A. B. Silva, J. G. Ramos, Á. R. Vasconcellos, R. Luzzi, Nonlinear higher-order hydrodynamics. Unification of kinetic and hydrodynamic approaches within a nonequilibrium statistical ensemble formalism, arXiv: 1210.7280.22, 2012.
29. C. G. Rodrigues, Á. R. Vasconcellos, R. Luzzi, Mesoscopic hydro-thermodynamics of phonons in semiconductors: heat transfer in III-nitrides, *Eur. Phys. J. B*, **86**:5, 200, (2013).
30. Á. R. Vasconcellos, A. R. B. de Castro, C. A. B. Silva, R. Luzzi, Mesoscopic hydro-thermodynamics of phonons, *AIP Adv*., **3**:7, 072106–072133 (2013).
31. C. A. B. Silva, C. G. Rodrigues, J. G. Ramos, R. Luzzi, Higher-order generalized hydrodynamics: foundations within a nonequilibrium statistical ensemble formalism, *Phys. Rev. E*, **91**:6, 063011, 15 pp. (2015)
32. C. G. Rodrigues, A. R. B. Castro, R. Luzzi, Higher-order generalized hydrodynamics of carriers and phonons in semiconductors in the presence of electric fields: macro to nano, *Phys. Stat. Sol. B*, **252**:12, 2802–2819 (2015).

33. C. G. Rodrigues, Á. R. Vasconcellos, R. Luzzi, Thermal conductivity in higher-order generalized hydrodynamics: characterization of nanowires of silicon and gallium nitride, *Phys. E*, **60**, 50–58 (2014).
34. D. N. Zubarev, *Non-equilibrium statistical thermodynamics*, Plenum-Consultants Bureau, New York, USA, 1974.
35. D. N. Zubarev, V. Morozov, and G. Röpke, *Statistical Mechanics of Non-equilibrium Processes: Basic Concepts, Kinetic Theory,* Akademie-Wiley VCH, Berlin, Germany, Vol. 1, 1996.
36. D. N. Zubarev, V. Morozov, and G. Röpke, *Statistical Mechanics of Non-equilibrium Processes: Relaxation and Hydrodynamic Processes,* Akademie-Wiley VCH, Berlin, Germany, Vol. 2, 1997.
37 D. N. Zubarev, The method of the non-equilibrium statistical operator and its application. I. The non-equilibrium statistical operator, *Fortschr. Physik*, **18**, 125–147 (1970).
38. D. N. Zubarev, in Reviews of Science and Technology: *Modern Problems of Mathematics*. Vol.**15**, pp. 131-226, (in Russian) ed. by R. B. Gamkrelidze, (Izd. Nauka, Moscow, 1980) [English Transl.: *J. Soviet Math*. **16**, 1509-1571 (1981)].
39. C. Gocke, G. Röpke, Master equation of the reduced statistical operator of an atom in a plasma, *Theoret. and Math. Phys*., 154:1, 26–51 (2008).
40. G. Röpke, *Nonequilibrium Statistical Physics*, Wiley-VCH, 2013.
41. G. Röpke, Nonequilibrium Statistical Operator. In: *Non-Equilibrium Particle Dynamics*; Kim, A.S., Ed.; Intech Open: London, UK, 2019; ISBN 978-1-83968-079-3, doi:10.5772/intechopen.84707.
42. P. P. Kostrobij, O. V. Viznovych, B. B. Markiv, M. V. Tokarchuk, Generalized kinetic equations for dense gases and liquids in the Zubarev nonequilibrium statistical operator method and Renyi statistics, *Theoret. and Math. Phys*., **184**:1, 1020–1032 (2015).
43. P. A. Glushak, B. B. Markiv, M. V. Tokarchuk, Zubarev's nonequilibrium statistical operator method in the generalized statistics of multiparticle systems, *Theoret. and Math. Phys.*, **194**:1, 57–73 (2018).
44. V. V. Ryazanov, Lifetime of System and Nonequilibrium Statistical Operator Method, *Fortschritte der Phusik/Progress of Physics*, v. **49**, N8-9, pp.885-893 (2001).
45. V. V. Ryazanov, First passage time and change of entropy, *Eur. Phys. J. B,* **94**, 242 (2021). https://doi.org/10.1140/epjb/s10051-021-00246-0.
46. V. V. Ryazanov, First-passage time: a conception leading to superstatistics. I. Superstatistics with discrete distributions. Preprint: *physics/0509098*, (2005); V. V. Ryazanov, First-passage time: a conception leading to superstatistics. II. Continuous distributions and their applications. Preprint: *physics/0509099*, (2005).
47. V. V. Ryazanov, S. G. Shpyrko, First-passage time: a conception leading to superstatistics. *Condensed Matter Physics*, **9**, 1(45), 71-80 (2006).
48. V. V. Ryazanov, Lifetime distributions in the methods of non-equilibrium statistical operator and superstatistics, *European Physical Journal B*, **72**, 629–639, (2009).
49. R. Metzler, G. Oshanin and S. Redner (ed), *First-Passage Phenomena and Their Applications*, Singapore: World Scientific, 2014, 608 p.
50. J. Masoliver, *Random Processes: First-Passage and Escape*, Singapore: World Scientific, *2018*, 388 p.
51. S B. Yuste, G. Oshanin, K. Lindenberg, O. Bénichou and J. Klafter, Survival probability of a particle in a sea of mobile traps: a tale of tails. *Phys. Rev. E* **78**, 021105 (2008).

52. É. Roldán, I. Neri, M. Dörpinghaus, H. Meyer, and F. Jülicher, Decision Making in the Arrow of Time. *Phys. Rev. Lett*., **115**, 250602 (2015).
53. K. Saito and A. Dhar, Waiting for rare entropic fluctuations. *Europhys. Lett*. **114**, 50004 (2016).
54. K. Ptaszynski, First-passage times in renewal and nonrenewal systems. *Phys. Rev. E*, **97**, 012127 (2018).
55. I. Neri, É. Roldán, and F. Jülicher, Statistics of Infima and Stopping Times of Entropy Production and Applications to Active Molecular Processes. *Phys. Rev. X*, **7**, 011019 (2017).
56. T. R. Gingrich and J. M. Horowitz, Fundamental Bounds on First Passage Time Fluctuations for Currents. *Phys. Rev. Lett*., **119**, 170601 (2017).
57. J. P. Garrahan, Simple bounds on fluctuations and uncertainty relations for first-passage times of counting observables. *Phys. Rev. E*, **95**, 032134 (2017).
58. P. Hänggi, P. Talkner, and M. Borkovec, Reaction-rate theory: fifty years after Kramers. *Rev. Mod. Phys*., **62**, 251-341 (1990).
59. A. Longtin, A. Bulsara, and F. Moss, Time-interval sequences in bistable systems and the noise-induced transmission of information by sensory neurons. *Phys. Rev. Lett*. **67**, 656-659 (1991).
60. H. C. Tuckwell, *Introduction to Theoretical Neurobiology*, v. 2, Cambridge University Press, Cambridge UK, 1988.
61. A. Molini, P. Talkner, G. G. Katul, A. Porporato, First passage time statistics of Brownian motion with purely time dependent drift and diffusion, *Physica A*, **390**, 1841–1852 (2011).
62. F. Bouchet and J. Reygner, Generalisation of the Eyring–Kramers Transition Rate Formula to Irreversible Diffusion Processes, *Ann. Henri Poincarè*, **17**, 3499–3532, (2016).
63. R. S. Maier and D. L. Stein, Limiting exit location distributions in the stochastic exit problem, *SIAM Journal on Applied Mathematics*, **57**, No. 3, 752-790 (1997).
64. J. Masoliver and J. Perellò, First-passage and escape problems in the Feller process, *Physical review E*, **86**, 041116 (2012).
65. M. V. Day, Recent progress on the small parameter exit problem, *Stochastics*, **20**, 121–150 (1987).
66. D. Hartich and A. Godec, Duality between relaxation and first passage in reversible Markov dynamics: rugged energy landscapes disentangled, *New J. Phys*, **20**, 112002 (2018).
67. D. Hartich and A. Godec, Interlacing relaxation and first-passage phenomena in reversible discrete and continuous space Markovian dynamics, *Journal of Statistical Mechanics: Theory and Experiment*, 2019 (2), 024002 (2019).
68. A. Godec and R. Metzler, Universal proximity effect in target search kinetics in the few-encounter limit, *Phys. Rev. X*, **6**, 041037, (2016).
69. V. V. Ryazanov, Nonequilibrium Thermodynamics based on the distributions containing lifetime as thermodynamic parameter, *Journal of thermodynamics*, Volume 2011, Article ID 203203, 10 pages, 2011. doi:10.1155/2011/203203.
70. V. V. Ryazanov, Nonequilibrium Thermodynamics and Distributions Time to achieve a Given Level of a Stochastic Process for Energy of System, *Journal of Thermodynamics*, vol. 2012, Article ID 318032, 5 pages, 2012. doi:10.1155/2012/318032.
71. I. Neri, Second Law of Thermodynamics at Stopping Times, *Phys. Rev. Lett*. **124**, 040601 (2020).
72. I. I. Gichman, A. V. Skorochod, *The theory of stochastic processes*, II, New-York, Springer-Verlag, 1974.

73. A. N. Shiryaev, *Statistical Sequential Analysis*, Amer. Mathematical Society, 1973, 174 p.
74. W. Feller, *An Introduction to Probability Theory and its Applications*, vol.2 (J. Wiley, New York, 1971).
75. Y. Klimontovich, *Statistical Physics*, Harwood-Academic Publishers, 734 p.
76. F. M. Kuni, *Statistical physics and thermodynamics*. Moskow, Nauka, 1981, 351 p.
77. R. Luzzi, A. R. Vasconcellos, J. G. Ramos, et al. Statistical Irreversible Thermodynamics in the Framework of Zubarev's Non-equilibrium Statistical Operator Method. *Theor. Math. Phys.,* **194** (2018), https://doi.org/10.1134/.
78. M. Abramowitz and I. A. Stegun, *Handbook of Mathematical Functions* (New York: Dover), 1965.
79. S. Sato and J. Inoue, Inverse gaussian distribution and its application, *Electronics and Communications in Japan* (Part III: Fundamental Electronic Science), **77**(1), 32-42, (1994).
80. G. Falasco and M. Esposito, Dissipation-Time Uncertainty Relation, *Phys. Rev. Lett.* **125**, 120604 (2020)
81. V. V. Ryazanov, Influence of Entropy Changes on First Passage Time in the Thermodynamics of trajectories. http://arxiv.org/abs/2303.07398 [cond-mat.stat-mech].
82. U. Seifert, From stochastic thermodynamics to thermodynamic inference, *Annual Review of Condensed Matter Physics*, vol. **10**, pp. 171-192, 2019.
83. Y. Hasegawa, Tan Van Vu, Fluctuation Theorem Uncertainty Relation, *Phys Rev Lett*, **123**(11):110602 (2019), doi: 10.1103/PhysRevLett.123.110602.
84. Arnab Pal, Shlomi Reuveni, Saar Rahav, Thermodynamic uncertainty relation for first-passage times on Markov chains, *Physical Review Research*, **3**(3) (2021).
85. R. L. Jack, J. P. Garrahan, and D. Chandler, Space-time thermodynamics and subsystem observables in a kinetically constrained model of glassy materials, *J. Chem. Phys.,* **125** (18), 184509, 2006.
86. J. P. Garrahan, R. L. Jack, V. Lecomte, E. Pitard, K. van Duijvendijk, and F. van Wijland, First-order dynamical phase transition in models of glasses: An approach based on ensembles of histories, *Journal of Physics A: Mathematical and Theoretical*, **42** (7), 075007, 2009.
87. L. O. Hedges, R. L. Jack, J. P. Garrahan, and D. Chandler, Dynamic order-disorder in atomistic models of structural glass formers, *Science*, **323**(5919):1309-13. doi: 10.1126/science.1166665, 2009.
88. J. P. Garrahan, R. L. Jack, V. Lecomte, E. Pitard, K. van Duijvendijk and F. van Wijland, Dynamic first-order phase transition in kinetically constrained models of glasses, *Phys. Rev. Lett*. **98**, 195702 (2007).
89. A. Budini, R. M. Turner, and J. P. Garrahan, Fluctuating observation time ensembles in the thermodynamics of trajectories, *Journal of Statistical Mechanics: Theory and Experiment*, **2014** (3), P03012.
90. J. P. Garrahan, Simple bounds on fluctuations and uncertainty relations for first-passage times of counting observables. *Phys. Rev. E,* **95**, 032134 (2017).
91. H. Touchette, The large deviation approach to statistical mechanics. Phys. Rep. 478, 1 (2009).
92. V. V. Ryazanov, First passage time of a given level and value of overjump for fluctuations of trajectory observables, http://arxiv.org/abs/2306.14664.
93. V. V. Ryazanov, Neutron Energy Distribution in a Nuclear Reactor Taking Account of the Finiteness of the Neutron Lifetime, *Atomic Energy*, **99,** 5, 782-786, November 2005.

94. V. V. Ryazanov, Influence of entropy changes on reactor period, http://arxiv.org/abs/2202.13349.

95. V.V.Ryazanov, Investigation of radiation-enhanced diffusion using first-passage time, *Radiation Physics and Chemistry*, Volume **203**, Part A, February 2023, 110608, http://arxiv.org/abs/2203.06449.

96. V. I. Tikhonov, M. A. Mironov, *Markov processes,* Moskow, Soviet Radio, 1977 (in Russian).

97. J. B. Madrid and S. D. Lawley, Competition between slow and fast regimes for extreme first passage times of diffusion, *Journal of Physics A: Mathematical and Theoretical*, **53**, Number 33, 500243 (2020).

98. A. Godec, R. Metzler, First passage time distribution in heterogeneity controlled kinetics: going beyond the mean first passage time, *Scientific Reports*, **6**, 20349 (2016).

99. S. G. Shpyrko, V. V. Ryazanov, Stochastic storage model and noise-induced phase transitions, *Eur. Phys J. B*, v.**54**, 2006, pp.345-354.

100. J. Kiukas and M. Gută, I. Lesanovsky and J. P. Garrahan, Equivalence of matrix product ensembles of trajectories in open quantum systems, *Physical Review E*, **92**(1), 2015, DOI:10.1103/PhysRevE.92.012132